# Proximity-induced superconductivity in a 2D Kondo lattice of an *f*-electron-based surface alloy

Howon Kim[1,*], Dirk K. Morr[2] and Roland Wiesendanger[1,*]

[1]Department of Physics, University of Hamburg, D-20355 Hamburg, Germany

[2]Department of Physics, University of Illinois at Chicago, Chicago, IL, 60607, USA

Corresponding authors(*): hkim@Physnet.uni-hamburg.de and wiesendanger@Physnet.uni-hamburg.de

**Realizing hybrids of low-dimensional Kondo lattices and superconducting substrates leads to fascinating platforms for studying the exciting physics of strongly correlated electron systems with induced superconducting pairing. Here, we report a scanning tunneling microscopy and spectroscopy study of a new type of two-dimensional (2D) La-Ce alloy grown epitaxially on a superconducting Re(0001) substrate. We observe the characteristic spectroscopic signature of a hybridization gap evidencing the coherent spin screening in the 2D Kondo lattice realized by the ultrathin La-Ce alloy film on normal conducting Re(0001). Upon lowering the temperature below the critical temperature of rhenium, a superconducting gap is induced with an in-gap Shiba band arising from the interaction of residual unscreened magnetic moments with the superconducting substrate. A positive correlation between the Kondo hybridization gap and the binding energy of the subgap Shiba band maximum is found. Our results open up a promising route toward the design of artificial superconducting Kondo and heavy fermion systems.**

Strongly correlated material systems have been an extremely active and exciting research area in condensed matter physics over the past decades [1,2]. Outstanding examples include heavy-fermion compounds [3] with extremely large effective electron masses exhibiting either superconductivity or magnetic order [4], high-$T_c$ cuprates [5], Kondo lattice systems [6], Kondo insulators [7], to mention only a few. Fascinating properties of strongly correlated electron systems range from unconventional superconductivity [8] to non-Fermi-liquid behavior, in particular close to quantum critical points [9]. Many of the unconventional superconductors are low-dimensional, layered materials close to a quantum phase transition. The interplay between different fundamental interactions results in complex phase diagrams and the emergence of novel exotic states of matter.

Most of the early pioneering work on strongly correlated electron systems has been performed based on bulk materials. The quality of the single crystals proved to be of great importance for revealing the intrinsic exciting physical properties of strongly correlated materials including cuprates and $f$-electron compounds [4]. Major challenges have been the purification of the starting elements as well as the microscopic material characterization, apart from bulk-sensitive methods such as X-ray diffraction. More recently, within the past two decades, investigations have focused more on low-dimensional systems, including ultrathin films [10–14], artificial 2D atomic arrays [15], and quasi-1D chains [16,17] revealing Kondo physics [18,19].

Remarkably, almost all previous investigations of atomic-scale Kondo systems using scanning tunneling microscopy and spectroscopy (STM/STS) techniques have been performed with transition metal-based magnetic impurities where the localized magnetic moments are due to $d$-electrons [20]. An exception is the early work on the Ce on Ag(111) system [21], where the authors assumed to study isolated Ce adatoms on the Ag substrate, but which later turned out to be Ce clusters [22]. Based on this particular sample system, namely Ce on Ag(111), it has been demonstrated that ordered 2D arrays of Ce adatoms can be achieved by self-assembly [23–25]. However, at the measurement temperature of around 4 K, neither a Kondo state of the individual Ce adatoms, nor the transition to the behavior of a 2D Kondo lattice could be observed.

Here, we report on the successful preparation of an ultrathin La-Ce alloy film on a clean Re(0001) substrate under ultra-high vacuum conditions. The choice of the La-Ce combination was motivated by the fact that the LaCe bulk material is a well-known Kondo alloy with a Kondo temperature below 1 K, which was already investigated intensively in the early seventies of the previous century [26,27]. The atomic-scale structure of the new type of 2D La-Ce alloy has been investigated by atomic-resolution STM, whereas the spatially resolved electronic properties have been revealed by STS measurements above and below the superconducting transition temperature ($T_{c,Re}$ ~ 1.6 K) of the Re(0001) substrate.

Tunneling spectroscopic data of the 2D La-Ce alloy on normal-conducting Re(0001) reveal a robust asymmetric double-peak resonance at the Fermi level, which is a hallmark of a Kondo lattice revealing a Kondo hybridization gap [28–36]. Below the superconducting transition temperature of the Re(0001) substrate, superconducting pairing is induced in the 2D La-Ce alloy via the proximity effect. The emergence of sub-gap excitations indicates the presence of magnetic bound states forming a low-energy Yu-Shiba-Rusinov (YSR) band [37–40]. A positive correlation between the Kondo hybridization gap and the binding energy of the YSR band maximum is found. Based on model calculations, we assign the observed sub-gap YSR band to the presence of residual magnetic moments due to only partial Kondo screening. The new type of 2D Kondo lattice proximitized to a superconductor provides novel insight into the mutual interactions between the two many-body effects and offers an exciting route toward the artificial design of low-dimensional superconducting heavy-fermion systems.

Figure 1(a) shows a representative STM topographic image of the ultrathin La-Ce alloy film grown on Re(0001). Ultra-thin La-Ce alloy films were prepared by a two-step *in situ* process under UHV conditions. Details of the epitaxial thin-film growth are described in the Supplemental Material [41]. The 2D La-Ce alloy starts to form at step edges of the pristine Re(0001) surface while parts of the Re(0001) terraces are covered with Ce clusters without a La wetting layer underneath. The 2D La-Ce alloy layer exhibits an ordered atomic lattice structure together with some randomly distributed local defects, being primarily Ce vacancy sites. A high-resolution zoomed-in STM image in Fig. 1(b) clearly shows the hexagonal lattice structure of the La-Ce alloy layer with a lattice constant of about 0.71 nm, which is equivalent to $\sqrt{7}\, a_{Re}$, where $a_{Re}$ is the lattice constant of the Re(0001) substrate ($a_{Re}$ = 0.274 nm). Interestingly, above the Ce vacancy sites of the La-Ce lattice, trimer-like structures formed by La atoms are visible in STM images obtained at low sample bias voltage [see Sec. S3 in [41]]. Based on the atomic-resolution STM data, an atomic structure model of the 2D La-Ce alloy can be derived, which is overlaid on the STM image of Fig. 1(b). Based on this model, the 2D La-Ce layer is composed of an ordered hexagonal array of La$_3$Ce complexes. Intermixing of lanthanides with the chosen Re(0001) substrate does not occur even during high-temperature annealing, in strong contrast to lanthanide (La, Ce, Gd) - based surface alloys formed on noble metal substrates, such as Au(111) [42,43].

To elucidate the electronic structure of the 2D La-Ce alloy in the normal-conducting state of the Re(0001) substrate, we obtained differential tunneling conductance (*dI/dV*) spectra at a temperature T above the superconducting transition temperature of the Re substrate ($T_{c,Re}$ ~1.6 K). Figure 2(a) shows tunneling spectra obtained at T = 1.7 K on a defect-free region of the 2D La-Ce alloy layer (red) as well as on a bare Re(0001) surface (grey). Within the energy range of ±0.04 eV, the La-Ce layer reveals an anomalous spectral feature around the Fermi energy ($E_F$), which is absent on the Re(0001) surface. In Fig. 2(b), the spatially averaged *dI/dV* spectrum within a smaller energy window shows a broad peak with a dip at E = -0.24 meV resulting in an asymmetric double-peak structure. This asymmetric double-peak resonance in the local density of states (LDOS) at $E=E_F$ is a characteristic signature of a Kondo lattice [30–32]. The peak-to-peak distance is determined to be 21.9 meV, which represents the magnitude of the hybridization gap ($\Delta_{hyb}$) in the LDOS for a Kondo lattice system.

In order to explore the spatial distribution of the spectral feature of the 2D La-Ce alloy layer, we show a spectroscopic line profile across the La-Ce layer including local defects (Fig. 2(c)), as shown in Fig. 2(d). A nearly uniform hybridization gap around $E_F$ is revealed with small spatial fluctuations caused by the presence of the local defect sites. For comparison, the averaged tunneling spectrum of Fig. 2(b) is plotted on the right-hand side of Fig 2(d). The nearly uniform spatial distribution of the hybridization gap shows that a coherent Kondo lattice state delocalized over the 2D La-Ce alloy film has formed.

Next, we investigated the evolution of the electronic structure of the La-Ce layer as superconductivity of the Re(0001) substrate is switched on. Fig. 3(a) shows the local tunneling spectra of the 2D La-Ce alloy obtained above (grey) and below $T_{c,Re}$ (blue) of Re. The spectrum below $T_{c,Re}$ reveals a proximity-induced superconducting gap at $E_F$ with asymmetric coherence peaks, while the hybridization gap due to the Kondo lattice state remains almost unaffected.

The latter is expected since the hybridization gap is significantly larger than the superconducting order parameter $\Delta_{Re}$ in the pure Re compound, such that the proximity induced superconducting gap is indeed only a small perturbation to the Kondo screened electronic structure. By zooming into a more narrow energy window, as shown in Fig. 3(b), a pronounced peak at $E_B$ = +0.21 meV and a peak with a fair intensity at $E_B$ = -0.37 meV [inset of Fig. 3(b)] are visible in the tunneling spectrum averaged over the ordered La-Ce alloy layer (red). Remarkably, both peaks are shifted to lower energy by $\Delta E = 80$ µeV with respect to the energetically symmetric coherence peaks at $E = \pm 0.29$ meV in the spectrum of bare superconducting Re(0001) (grey). While the absence of symmetrically located coherence peaks is unexpected, the presence of an in-gap state (with respect to the bare Re(0001) surface) at +0.21 meV provides the first indication that the uniformly shifted coherence peaks are related to the presence of residual magnetism, and the formation of a YSR band [38,40].

To further elucidate the spatial distribution of the sub-gap states in the superconducting 2D La-Ce alloy, we analyze spatially-resolved tunneling spectra across the La-Ce layer on Re(0001). The plot on the right-hand side of Fig. 3(d) shows a spectroscopic line profile along the dotted line in the STM image of Fig. 3(c). The observed pronounced and spatially homogeneous peak at around $E = +0.21$ meV inside the superconducting gap indicates that this peak is not related to some local defect state, but represents a coherent feature of the Kondo lattice in which superconductivity is proximity-induced. For comparison, an averaged spectrum is shown in the middle of Fig. 3(d), which is clearly distinct from the one measured above an uncovered Re(0001) area (left part of Fig. 3(d)).

To understand the experimentally observed uniform shift of the Re coherence peaks to lower energies in the La-Ce alloy, we note that the observed Kondo resonance in the larger energy window most likely arises from the coupling of the localized $f$ orbital-derived magnetic moments of Ce with the itinerant conduction electrons, and the concomitant coherent Kondo screening of the Ce moments. While it is presently unclear whether the Ce moments are fully screened above $T_c$, the opening of a superconducting gap below $T_c$, and the concomitant gapping of the conduction band's low-energy degrees of freedom, is expected to lead to a partial unscreening of the Ce magnetic moments. The unscreened portion of the local moments, denoted by $S$, can interact with the conduction electrons via the Kondo coupling $J$, and/or with the $f$-electron states via a Heisenberg exchange $I$. The largest effect of this interaction will occur for states near the Fermi energy, which in the generic mean-field approximation of the Kondo lattice model, are predominantly of $f$-electron character [see supplementary Figure S1(a)]. It is therefore likely that the main effect of the partially unscreened magnetic moment are uniform energy shifts of $\pm IS$ in the electronic structure of the two spin species of the $f$-electrons, corresponding to the formation of YSR bands of predominant $f$-electron character on the background of the partially Kondo screened and superconducting La-Ce/Re hybrid system. The fact that the experimental results reveal only a uniform downward shift of the electronic structure to negative energies could imply that the STM tip itself is spin-polarized, possibly by picking up Ce atoms, thus leading to a preferential tunneling of electrons of only one spin-polarization. To demonstrate this effect, we consider a large-N theory for the Kondo lattice, using a generic two-band model, with proximity-induced superconductivity, and a residual unscreened moment that interaction with the heavy $f$-electron states via a Heisenberg exchange $I$ (for details, see Sec. S2 in [41]) The resulting theoretical LDOS in the normal state above $T_c$ shown in Fig. 3(e) reproduces all salient features of the experimental $dI/dV$ in Fig. 3(a): a hybridization gap of about 20 meV, a slightly asymmetric Kondo resonance, and a minimum in the LDOS around zero energy. Below $T_c$, a superconducting gap opens on the background of the Kondo resonance. A comparison of the low-energy LDOS below $T_c$ in the presence and absence of the Kondo lattice, as shown in Fig. 3(f), visualizes an asymmetric shift in the LDOS for the Kondo lattice. This asymmetry is a direct signature of the presence of unscreened moments, and the predominant tunneling into one of the spin channels due to the likely spin polarization of the tip, as discussed above.

Finally, we investigate the dependence of the binding energy $E_B$ of the sub-gap YSR band on the magnitude of the hybridization gap $\Delta_{hyb}$. As described above, despite nearly uniform spatial distributions of both $\Delta_{hyb}$ and $E_B$, we observe sizable variations near local defects, which are likely to induce an inhomogeneous distribution of unscreened magnetic moments. Fig. 4(a) presents a correlation map of ($\Delta_{hyb}$, $E_B$) extracted from local tunneling spectra above the ordered 2D La-Ce alloy (red) and above edge- or defect sites (grey). Interestingly, we observe a positive correlation between $\Delta_{hyb}$ and $E_B$ for the ordered 2D La-Ce alloy as indicated by the red ellipse, i.e. $E_B$ moves to the superconducting gap edge for larger $\Delta_{hyb}$. On the other hand, no correlation is found for the edge- and defect-sites of the La-Ce layer. To demonstrate the positive correlation for the ordered La-Ce alloy based on primary STS data, we present two sets of tunneling spectra obtained at three different spots (P1 - P3) above the ordered 2D La-Ce alloy, for extracting $\Delta_{hyb}$ and $E_B$ in Fig. 4(b) and 4(c), respectively. As $\Delta_{hyb}$ is increased from P1 to P3 (Fig. 4(b)), $E_B$ is gradually shifted toward the gap edge (Fig. 4(c)).

For a single magnetic impurity interacting with a superconductor, the relationship between the Kondo coupling $T_K$ and the binding energy $E_{YSR}$ of the YSR state has previously been discussed intensively. According to Matsuura's model based on the local Fermi-liquid approach for the strong Kondo regime ($\Delta<<k_B T_K$) [44] which is realized in our experiment, the energy of the YSR bound states was found to be:

$$E_{YSR} = \Delta \frac{1-\alpha^2}{1+\alpha^2}, \quad \text{where} \quad \alpha \sim \frac{\pi\Delta}{4k_B T_K} \ln\left(\frac{4k_B T_K}{\pi\Delta} e\right) .$$

For a Kondo lattice system, $\Delta_{\text{hyb}}$ characterizes the strength of the Kondo coupling ($\Delta_{\text{hyb}}=2k_B T_K$). Large values of $\Delta_{\text{hyb}}$ imply strong Kondo screening. Thus with increasing $\Delta_{\text{hyb}}$, and hence $T_K$, $\alpha$ decreases and $E_{YSR}$ moves closer to the gap edge, which is qualitatively consistent with the positive correlation between $\Delta_{\text{hyb}}$ and $E_B$ observed in our experiment.

In summary, we have explored a novel type of 2D La-Ce alloy epitaxially grown on a Re(0001) substrate. Our detailed STS investigations have revealed clear spectroscopic signatures of a Kondo lattice state above and below $T_c$ of Re as well as the formation of a sub-gap YSR band of the superconducting hybrid system. The origin of the sub-gap YSR band has been attributed to the presence of partially unscreened magnetic moments even in the strong-coupling Kondo regime ($T_c \ll T_K$). Additionally, we have shown that the spatial variation of the hybridization gap indicates local variations of the effective Kondo screening. This in turn leads to spatial variations of the residual magnetic moments resulting in the observed variation of the sub-gap YSR states. Our results provide novel insight into the behavior of low-dimensional Kondo lattice – superconductor hybrid systems, which can become a versatile platform for studying microscopic aspects of exotic quantum states in artificially designed superconducting $f$-electron based materials.

**Figure 1.**

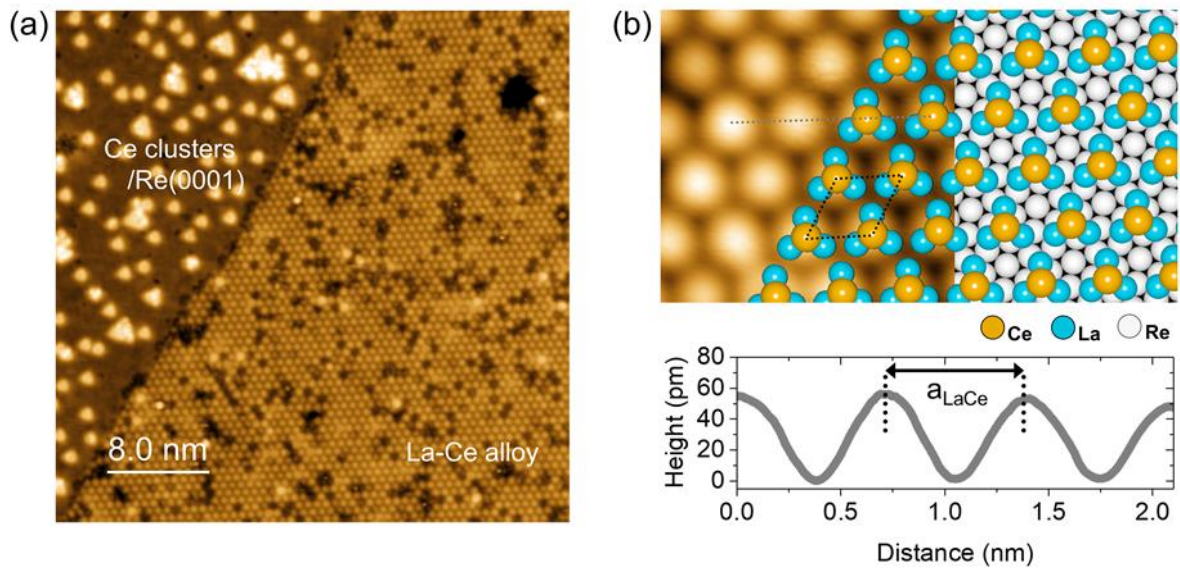

**Figure 1. Formation of a 2D La-Ce alloy on a Re(0001) substrate.** (a) Constant-current STM topography image (40 × 40 nm$^2$) showing the formation of an ultrathin La-Ce alloy layer on a Re(0001) substrate (right part). Without La, the Ce forms clusters on the Re(0001) surface (left part). (b) (top) A zoomed-in STM image revealing the atomic-scale structure of the 2D La-Ce alloy with an atomic ball model superimposed. The unit cell of the La-Ce lattice structure is highlighted by a dotted rhombus. (bottom) Surface profile along the dotted gray line in (b). The lattice constant ($a_{LaCe}$) is 0.71 nm which corresponds to $\sqrt{7}a_{Re}$, where $a_{Re}$=0.27 nm is the lattice constant of the Re(0001) substrate. Tunneling current: $I_T$=1.0 nA; applied sample bias voltage: $V_S$=+40 mV.

**Figure 2.**

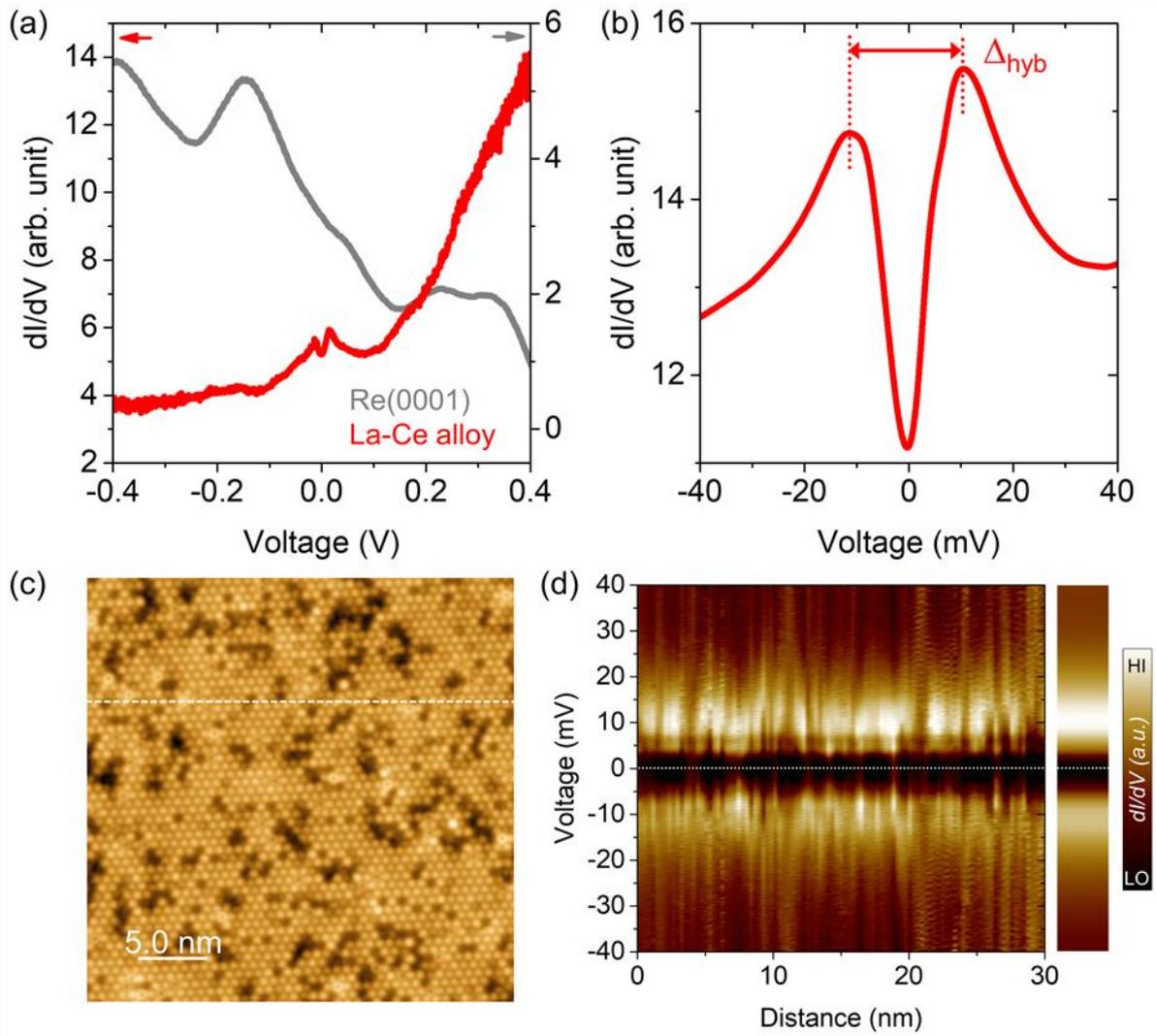

**Figure 2. Spectroscopic signatures of the 2D La-Ce alloy on Re(0001) in the normal conducting state.** (a) Representative local tunneling spectra taken on the ultrathin La-Ce alloy layer (red, 1.0 nA/-400 mV) and on the uncovered, clean Re(0001) surface (gray, 1.0 nA/+600 mV). (b) Local tunneling spectrum averaged over an extended area of the 2D La-Ce layer revealing an asymmetric double-peak feature around the Fermi level. The peak-to-peak distance corresponding to the hybridization gap $\Delta_{hyb}$ is 21.9 meV, and the dip-center is located at energy E= -0.24 meV. (c) Constant-current STM topography image (1.0 nA/50 mV, 30 × 30 nm$^2$) of the La-Ce alloy layer corresponding to the location where the tunneling spectroscopic data has been obtained. (d) Tunneling spectroscopic map obtained on the 2D La-Ce layer along the dashed line in (c) revealing a robust hybridization gap $\Delta_{hyb}$ over the entire area. For comparison, the spectrum in (b) is plotted in the same color-scale (right side). All STM and STS data shown is this figure were obtained at a sample temperature T=1.7 K in the normal conducting state of the Re(0001) substrate.

**Figure 3.**

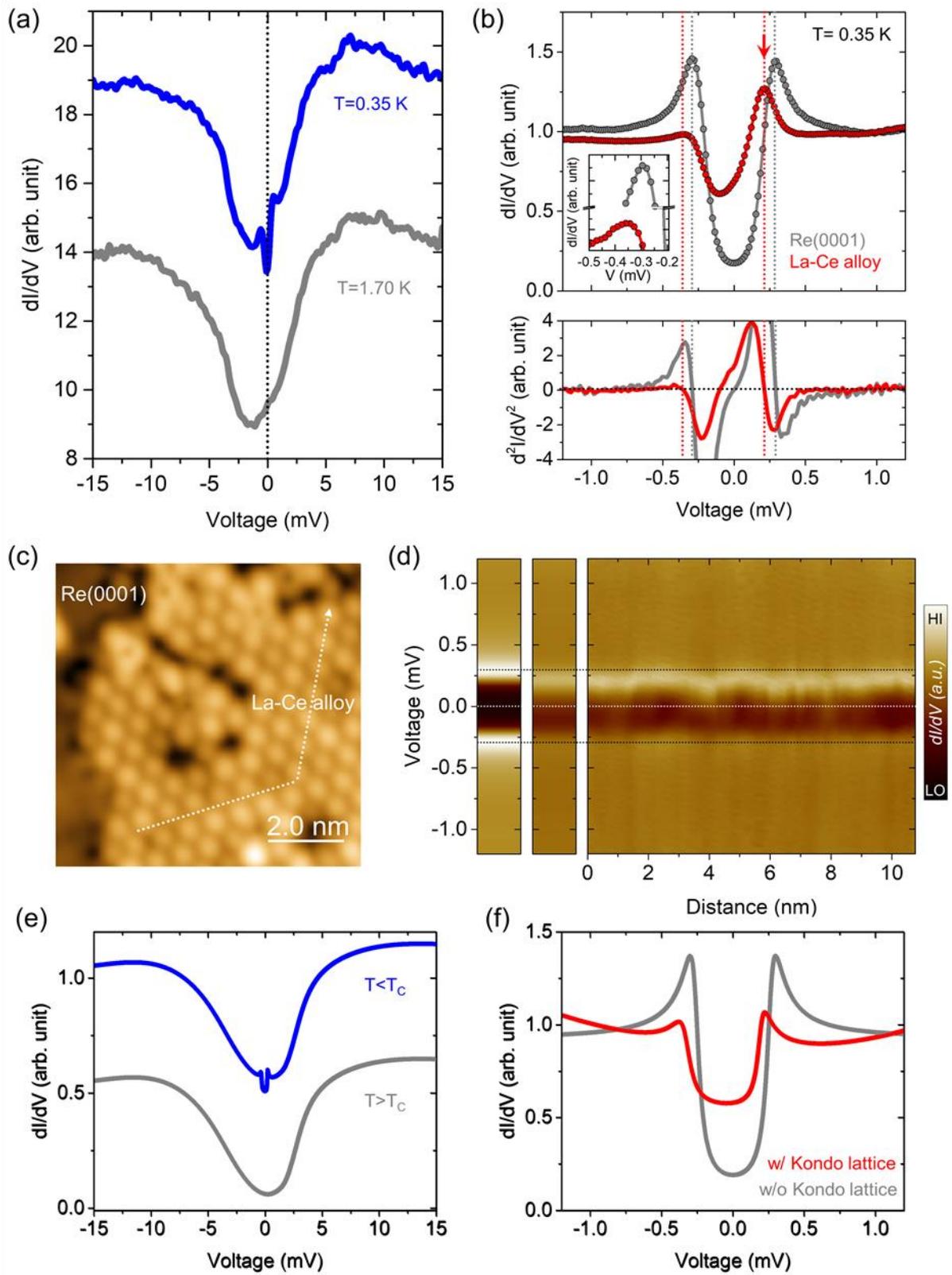

**Figure 3. Spectroscopic signatures of the 2D La-Ce alloy on Re(0001) in the superconducting state.** (a) Local tunneling spectrum (blue) obtained on the 2D La-Ce alloy layer at a temperature of 0.35 K, i.e. below $T_{c,Re}$. For comparison, the tunneling spectrum for the same energy range is shown at a temperature of 1.7 K, i.e. above $T_{c,Re}$. The superconductivity-induced spectral feature is visible around $E_F$ when $T$ is below $T_{c,Re}$. (b) Spatially averaged, low-energy tunneling spectra measured on the La-Ce alloy layer (red) and on the uncovered Re(0001) surface near a La-Ce layer (gray). A pronounced peak inside the superconducting gap of Re appears at E = +0.22 meV (red arrow). (Inset) A zoom-in focusing on the small energy range and corresponding spectral shape around the coherence peak at negative bias voltage. (Bottom) Numerical differentiation of the spectra clearly indicating the energy positions of the peaks in the $dI/dV$ spectra at $d^2I/dV^2 = 0$ (gray- and red-dotted lines for the spectra on the La-Ce alloy and Re(0001) surface, respectively). (c) Constant-current STM topographic image of a La-Ce alloy region (right part) next to the bare Re(0001) surface (left part). (d) Tunneling spectroscopic map obtained on the La-Ce alloy region along the dotted line in (c) (right). For direct comparison, the spatially averaged tunneling spectrum for the La-Ce alloy layer (middle) and the bare superconducting Re(0001) substrate (left) are plotted using the same color-scale. (e) Theoretical LDOS above (grey) and below (blue) $T_c$. (f) Low-energy LDOS below $T_c$ in the presence (red) and absence (grey) of the Kondo lattice. For details of the theoretical model, see Sec. 2 in [41].

**Figure 4.**

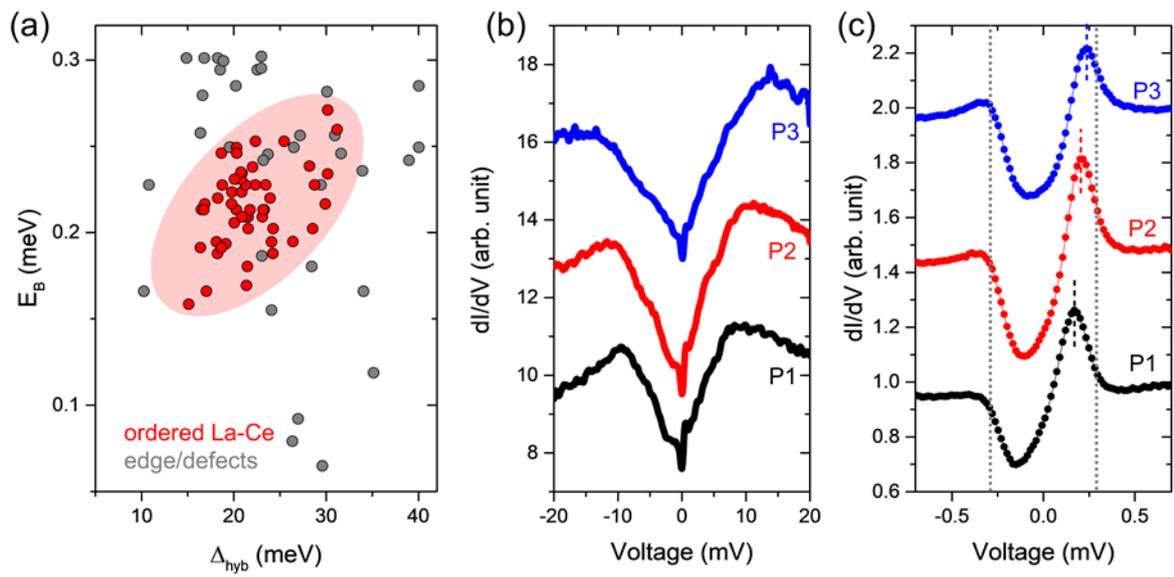

**Figure 4. Correlation between hybridization gap and Shiba band maximum of the 2D Kondo lattice – superconductor hybrid system.** (a) Correlation map showing the Shiba band maximum $E_B$ as a function of the hybridization gap $\Delta_{hyb}$ extracted from local tunneling spectra obtained on the ordered La-Ce alloy layer (red) and edge/defect-sites of the La-Ce layer (grey). A positive correlation is apparent as marked with an ellipse to guide the eyes. (b) and (c) Representative local tunneling spectra obtained at three different spatial locations (P1 to P3) with a larger energy window to extract $\Delta_{hyb}$ (b), and a smaller one to extract $E_B$ (c). Colored dashed lines indicate the values for $E_B$, whereas the gap edges are depicted with grey dotted lines at E=0.29 meV. All tunneling spectra were obtained at T=0.35 K. Tunneling parameters: (b) $I_T$= 1.0 nA, $V_S$= 20 mV, and $V_{ac}$=0.3 mV$_{rms}$; (c) $I_T$= 0.8 nA, $V_S$= +1.2 mV, and $V_{ac}$=0.03 mV$_{rms}$.

*Supplementary Material*

See Supplemental Material at the LINK (the corresponding link will be provided by the publisher) for the experimental and theoretical methods and supplemental discussions including Refs. [6,15,45–49].

*Acknowledgement*

We thank L. Schneider, J. Wiebe, A. Belozerov, R. Mozara and T. Wehling for fruitful discussions and D. Schreyer for experimental support at the initial stage of this work. H.K. and R.W. acknowledge financial support from the European Research Council via project No. 786020 (ERC Advanced Grant ADMIRE). D.K.M was supported by the U.S. Department of Energy, Office of Science, Basic Energy Sciences, under Award No. DE-FG02-05ER46225.

*References*

[1] P. Fulde, *Electron Correlations in Molecules and Solids* (Springer Berlin Heidelberg, Berlin, Heidelberg, 1995).

[2] S. Paschen and Q. Si, Nat. Rev. Phys. **3**, 9 (2020).

[3] G. R. Stewart, Rev. Mod. Phys. **56**, 755 (1984).

[4] C. Pfleiderer, Rev. Mod. Phys. **81**, 1551 (2009).

[5] E. Dagotto, Rev. Mod. Phys. **66**, 763 (1994).

[6] D. K. Morr, Rep. Prog. Phys. **80**, 014502 (2017).

[7] G. Aeppli and Z. Fisk, Comments Cond. Mat. Phys. **16**, 155 (1992).

[8] G. R. Stewart, Adv. Phys. **66**, 75 (2017).

[9] H. V. Löhneysen, A. Rosch, M. Vojta, and P. Wölfle, Rev. Mod. Phys. **79**, 1015 (2007).

[10] N. Tsukahara, S. Shiraki, S. Itou, N. Ohta, N. Takagi, and M. Kawai, Phys. Rev. Lett. **106**, 187201 (2011).

[11] P. Coleman, Science **327**, 969 (2010).

[12] H. Shishido, T. Shibauchi, K. Yasu, T. Kato, H. Kontani, T. Terashima, and Y. Matsuda, Science **327**, 980 (2010).

[13] M. Shimozawa, S. K. Goh, T. Shibauchi, and Y. Matsuda, Rep. Prog. Phys. **79**, 074503 (2016).

[14] K. Ienaga, S. Kim, T. Miyamachi, and F. Komori, Phys. Rev. B **104**, 165419 (2021).

[15] J. Figgins, L. S. Mattos, W. Mar, Y.-T. Chen, H. C. Manoharan, and D. K. Morr, Nat. Commun. **10**, 5588 (2019).

[16] A. DiLullo, S.-H. Chang, N. Baadji, K. Clark, J.-P. Klöckner, M.-H. Prosenc, S. Sanvito, R. Wiesendanger, G. Hoffmann, and S.-W. Hla, Nano Lett. **12**, 3174 (2012).

[17] D.-J. Choi, R. Robles, S. Yan, J. A. J. Burgess, S. Rolf-Pissarczyk, J.-P. Gauyacq, N. Lorente, M. Ternes, and S. Loth, Nano Lett. **17**, 6203 (2017).

[18] J. Kondo, Prog. Theor. Phys. **32**, 37 (1964).

[19] A. C. Hewson, *The Kondo Problem to Heavy Fermions* (Cambridge University Press, Cambridge, 1993).

[20] V. Madhavan, W. Chen, T. Jamneala, M. F. Crommie, and N. S. Wingreen, Science **280**, 567 (1998).

[21] J. Li, W.-D. Schneider, R. Berndt, and B. Delley, Phys. Rev. Lett. **80**, 2893 (1998).

[22] M. Ternes, A. J. Heinrich, and W.-D. Schneider, J. Phys. Condens. Matter **21**, 053001 (2009).

[23] F. Silly, M. Pivetta, M. Ternes, F. Patthey, J. P. Pelz, and W.-D. Schneider, Phys. Rev. Lett. **92**, 016101 (2004).

[24] M. Ternes, C. Weber, M. Pivetta, F. Patthey, J. P. Pelz, T. Giamarchi, F. Mila, and W.-D. Schneider, Phys. Rev. Lett. **93**, 146805 (2004).

[25] F. Silly, M. Pivetta, M. Ternes, F. Patthey, J. P. Pelz, and W.-D. Schneider, New J. Phys. **6**, 16 (2004).

[26] E. Umlauf, J. Schneider, R. Meier, and H. Kreuzer, J. Low Temp. Phys. **5**, 191 (1971).

[27] P. M. Chaikin and T. W. Mihalisin, Phys. Rev. B **6**, 839 (1972).

[28] C. M. Varma, Rev. Mod. Phys. **48**, 219 (1976).

[29] P. Coleman, in *Handbook of Magnetism and Advanced Magnetic Materials* (John Wiley & Sons, Ltd, Chichester, UK, 2007).

[30] M. Maltseva, M. Dzero, and P. Coleman, Phys. Rev. Lett. **103**, 206402 (2009).

[31] J. Figgins and D. K. Morr, Phys. Rev. Lett. **104**, 187202 (2010).

[32] P. Wölfle, Y. Dubi, and A. V. Balatsky, Phys. Rev. Lett. **105**, 246401 (2010).

[33] A. R. Schmidt, M. H. Hamidian, P. Wahl, F. Meier, A. V. Balatsky, J. D. Garrett, T. J. Williams, G. M. Luke, and J. C. Davis, Nature **465**, 570 (2010).

[34] P. Aynajian, E. H. da Silva Neto, C. V. Parker, Y. Huang, A. Pasupathy, J. Mydosh, and A. Yazdani, Proc. Natl. Acad. Sci. **107**, 10383 (2010).

[35] P. Aynajian, E. H. da Silva Neto, A. Gyenis, R. E. Baumbach, J. D. Thompson, Z. Fisk, E. D. Bauer, and A. Yazdani, Nature **486**, 201 (2012).

[36] S. Rößler, T.-H. Jang, D.-J. Kim, L. H. Tjeng, Z. Fisk, F. Steglich, and S. Wirth, Proc. Natl. Acad. Sci. **111**, 4798 (2014).

[37] Yu Luh, Acta Phys. Sin. **21**, 75 (1965).

[38] H. Shiba, Prog. Theor. Phys. **40**, 435 (1968).

[39] A. I. Rusinov, Sov. J. Exp. Theor. Phys. **29**, 1101 (1969).

[40] A. V. Balatsky, I. Vekhter, and J.-X. Zhu, Rev. Mod. Phys. **78**, 373 (2006).

[41] See Supplemental Material at the LINK (the corresponding link will be provided by the publisher) for the experimental and theoretical methods and supplemental discussions including Refs. [6,15,45–49].

[42] M. Ormaza, L. Fernández, S. Lafuente, M. Corso, F. Schiller, B. Xu, M. Diakhate, M. J. Verstraete, and J. E. Ortega, Phys. Rev. B **88**, 125405 (2013).

[43] M. Bazarnik, M. Abadia, J. Brede, M. Hermanowicz, E. Sierda, M. Elsebach, T. Hänke, and R. Wiesendanger, Phys. Rev. B **99**, 174419 (2019).

[44] T. Matsuura, Prog. Theor. Phys. **57**, 1823 (1977).

[45] S. Ouazi, T. Pohlmann, A. Kubetzka, K. von Bergmann, and R. Wiesendanger, Surf. Sci. **630**, 280 (2014).

[46] P. Coleman, Phys. Rev. B **29**, 3035 (1984).

[47] D. M. Newns and N. Read, Adv. Phys. **36**, 799 (1987).

[48] G. Kotliar and J. Liu, Phys. Rev. B **38**, 5142 (1988).

[49] A. Palacio-Morales, E. Mascot, S. Cocklin, H. Kim, S. Rachel, D. K. Morr, and R. Wiesendanger, Sci. Adv. **5**, eaav6600 (2019).

# Supplemental Material for

# Proximity-induced superconductivity in a 2D Kondo lattice of an *f*-electron-based surface alloy

Howon Kim[1,*], Dirk K. Morr[2] and Roland Wiesendanger[1,*]

Corresponding authors(*): hkim@Physnet.uni-hamburg.de and wiesendanger@Physnet.uni-hamburg.de

## Section 1. Experimental methods

### A) Sample and STM tip preparation

The Re(0001) single crystal used as a superconducting substrate in this work was prepared by repeated cycles of $O_2$ annealing at 1400 K followed by flashing at 1800 K to obtain an atomically flat Re(0001) surface [1]. Lanthanum and cerium were deposited separately *in situ* under ultrahigh vacuum conditions by electron beam evaporation of pure La pieces (99.9+%, MaTeck, Germany) and a Ce rod (99.9+%, MaTeck, Germany) from molybdenum crucibles. The deposition rates of both materials were calibrated separately by depositing them onto a clean Re(0001) surface prior to the preparation of the ultrathin La-Ce alloy layer. The 2D La-Ce alloy was prepared by a two-step *in situ* process. Initially, a monolayer of La was deposited onto the Re(0001) substrate followed by annealing at 950 K for 5 minutes, resulting in the formation of a La wetting layer on the Re(0001) surface. Subsequently, a sub-monolayer coverage of Ce was deposited onto the La wetting layer followed by annealing at 950K for 15 minutes. Finally, the samples were transferred into the cryostat without breaking vacuum. Commercially available Pt-Ir tips were used as STM probes being sharpened and cleaned by *in situ* tip treatments.

### B) STM/STS measurements

All STM/STS experiments were performed in a $^3$He-cooled low-temperature STM system (USM-1300S, Unisoku, Japan) operating at T = 0.35 K up to 1.70 K under ultra-high vacuum conditions. Tunneling spectra were obtained by measuring the differential tunneling conductance (*dI/dV*) using a standard lock-in technique under opened feedback loop with a frequency of 1128 Hz and modulation voltages of 0.03 mV$_{rms}$ and 0.30 mV$_{rms}$ in order to resolve the spectral features related with superconductivity and Kondo hybridization, respectively. The bias voltage was applied to the sample and the tunneling current was measured through the tip using a commercially available controller (Nanonis, SPECS).

**Section 2. Theoretical model**

To compute the electronic and spectroscopic structure of the Kondo screened magnetic layer above $T_c$, our starting point can either be a Hamiltonian in the band basis, or one in the orbital basis. Since the physical properties near the Fermi energy are dominated by the heavy magnetic states, we show below that both starting points lead to an essentially identical low-energy structure. For simplicity, we therefore start with a Hamiltonian that yields a description of the low-energy physics in terms of a band basis given by [2]

$$H = \sum_{\mathbf{k},\sigma} \varepsilon_\mathbf{k} c^\dagger_{\mathbf{k}\sigma} c_{\mathbf{k}\sigma} + \sum_{\mathbf{k},\sigma} \chi_\mathbf{k} f^\dagger_{\mathbf{k}\sigma} f_{\mathbf{k}\sigma} + V \sum_{\mathbf{k},\sigma} (f^\dagger_{\mathbf{k}\sigma} c_{\mathbf{k}\sigma} + c^\dagger_{\mathbf{k}\sigma} f_{\mathbf{k}\sigma}) \quad (1)$$

where

$$\varepsilon_\mathbf{k} = -2 t_c [\cos(k_x) + \cos(k_y)] - \mu_c \quad (2)$$

$$\chi_k = -2 t_f [\cos(k_x) + \cos(k_y)] - \mu_f \quad (3)$$

are the dispersions of the light conduction electrons ($\varepsilon_\mathbf{k}$) and heavy magnetic states ($\chi_k$), and $V$ is the hybridization between them. Here, $c^\dagger_{\mathbf{k}\sigma}$ ($f^\dagger_{\mathbf{k}\sigma}$) creates an electron with momentum **k** and sin $\sigma$ in the light (heavy magnetic) bands. This Hamiltonian is obtained from the $U \to \infty$ limit of the Anderson Kondo lattice Hamiltonian [3–5]. Note that the dispersion of the heavy magnetic states (also referred to as the $f$-electron states) in general arises from a Heisenberg exchange coupling $I$ between neighboring magnetic atoms [2]. Diagonalizing the above Hamiltonian yields

$$H = \sum_{\mathbf{k},\sigma} \left( E^\alpha_\mathbf{k} \alpha^\dagger_{\mathbf{k}\sigma} \alpha_{\mathbf{k}\sigma} + E^\beta_\mathbf{k} \beta^\dagger_{\mathbf{k}\sigma} \beta_{\mathbf{k}\sigma} \right) \quad (4)$$

where

$$E_{\mathbf{k}}^{\alpha,\beta} = \frac{\varepsilon_{\mathbf{k}} + \chi_{\mathbf{k}}}{2} \pm \sqrt{\left(\frac{\varepsilon_{\mathbf{k}} - \chi_{\mathbf{k}}}{2}\right)^2 + V^2} \qquad (5)$$

For the theoretical results shown in the main text, we used the following parameters:

$t_c = 100\ meV, \mu = -3.618 t_c, t_f = 0.025 t_c, \mu_f = 0.07 t_c, V = 0.3 t_c$, yielding the dispersion shown in Fig. S1(a) in the normal state. We assume that superconductivity is proximity induced in the La-Ce layer on the background of these two bands, which are formed at temperatures much higher than $T_c$. In general, only one of the two bands, $E_{\mathbf{k}}^{\beta}$, crosses the Fermi surface [see Fig. S1(a)], and we thus assume that only this band becomes superconducting via proximity coupling to the Re system below $T_c$, Moreover, since the electronic states near the Fermi surface are predominantly of heavy $f$-electron character, as follows from Fig. S1(a), the main effect of the unscreened $f$-electron magnetic moment arises from its coupling to heavy $f$-electron states via the Heisenberg exchange coupling $I$. Thus, only the bands $E_{\mathbf{k}}^{\beta}$ crossing the Fermi surface will be subject to an effective Zeeman splitting by $\pm IS$, where $S$ is the magnitude of the unscreened moment. The Hamiltonian below $T_c$ then takes the form

$$H = \sum_{\mathbf{k},\sigma} E_{\mathbf{k}}^{\alpha} \alpha_{\mathbf{k}\sigma}^{\dagger} \alpha_{\mathbf{k}\sigma} + \sum_{\mathbf{k},\sigma}\left(E_{\mathbf{k}}^{\beta} - IS\,\text{sgn}\sigma\right)\beta_{\mathbf{k}\sigma}^{\dagger}\beta_{\mathbf{k}\sigma} + \sum_{\mathbf{k}} \Delta\left(\beta_{\mathbf{k}\uparrow}^{\dagger}\beta_{-\mathbf{k}\downarrow}^{\dagger} + \beta_{-\mathbf{k}\downarrow}\beta_{\mathbf{k}\uparrow}\right) \qquad (6)$$

where $\Delta$ is the proximity induced superconducting order parameter. The resulting dispersion in the superconducting state is shown in Fig. S1(b), and a zoom-in is presented in Fig. S1(c). A comparison of Figs. S1 (a) and (c) reveals that the superconducting gap opens in the predominantly $f$-electron part of the dispersion, since the $c$-electron dispersion is pushed away from the Fermi energy due to the opening of the hybridization gap. The resulting retarded Greens function for the spin-↑ $c$-electrons is then given by

$$G_{cc}(k,\uparrow,\omega) = \frac{u_k^2}{\omega + i\Gamma(\omega) - E_{\mathbf{k}}^{\alpha}} + v_k^2 \frac{\omega + IS + i\Gamma(\omega) - E_{\mathbf{k}}^{\beta}}{\left(\omega + IS + i\Gamma(\omega)\right)^2 - \left(\Omega_{\mathbf{k}}^{\beta}\right)^2} \qquad (7)$$

where

$$u_k^2, v_k^2 = \frac{1}{2}\left(1 \pm \frac{\frac{\varepsilon_{\mathbf{k}} - \chi_{\mathbf{k}}}{2}}{\sqrt{\left(\frac{\varepsilon_{\mathbf{k}} - \chi_{\mathbf{k}}}{2}\right)^2 + V^2}}\right) \qquad (8)$$

and

$$\Omega_{\mathbf{k}}^{\beta} = \sqrt{(E_{\mathbf{k}}^{\alpha})^2 + \Delta^2} \tag{9}$$

To model the experimentally observed broadening in the differential conductance $dI/dV$, we consider a frequency dependent inverse particle lifetime given by

$$\Gamma(\omega) = \Gamma_0 + \alpha \omega^2 \tag{10}$$

with $\Gamma_0 = 0.03 meV$ and $\alpha = 0.238/meV$.

The local density of states (DOS) for the translationally invariant system is then given by

$$N_{cc}(\uparrow,\omega) = -\frac{1}{\pi} \text{Im}[G_{cc}(\mathbf{r},\uparrow,\omega)] \tag{11}$$

where

$$G_{cc}(\mathbf{r},\uparrow,\omega) = \int \frac{d^2k}{(2\pi)^2} G_{cc}(\mathbf{k},\uparrow,\omega) \tag{12}$$

An alternative approach, which introduces the proximity induced superconducting order parameter in the orbital basis, starts from the Hamiltonian

$$\begin{aligned} H = & \sum_{\mathbf{k},\sigma} (\varepsilon_{\mathbf{k}} - JS\text{sgn}\sigma) c_{\mathbf{k}\sigma}^{\dagger} c_{\mathbf{k}\sigma} + \sum_{\mathbf{k},\sigma} (\chi_{\mathbf{k}} - IS\text{sgn}\sigma) f_{\mathbf{k}\sigma}^{\dagger} f_{\mathbf{k}\sigma} \\ & + V \sum_{\mathbf{k},\sigma} (f_{\mathbf{k}\sigma}^{\dagger} c_{\mathbf{k}\sigma} + c_{\mathbf{k}\sigma}^{\dagger} f_{\mathbf{k}\sigma}) + \sum_{\mathbf{k}} (\Delta_c c_{\mathbf{k}\uparrow}^{\dagger} c_{-\mathbf{k}\downarrow}^{\dagger} + \Delta_f f_{\mathbf{k}\uparrow}^{\dagger} f_{-\mathbf{k}\downarrow}^{\dagger} + h.c.) \end{aligned} \tag{13}$$

Here, we assume that in addition to the shift in the $f$-electron dispersion by $\pm IS$ arising from the magnetic Heisenberg exchange $I$ between neighboring magnetic moments, the unscreened moment also couples to the light band via the Kondo coupling $J$. However, since the light states are shifted away from the Fermi energy due to the hybridization with the heavy $f$-electron states, the effect of the Zeeman splitting $\pm JS$ on the low energy density of states is negligible Moreover, since $V \gg \Delta_{c,f}$ (reflecting the experimental observation that the hybridization gap is much larger than the superconducting gap), the low-energy density of states is predominantly determined by the $f$-electron states, explaining why the Hamiltonians in Eqs.(6) and (13) yield essentially identical results for the low energy density of states. Finally, we note that $\Delta_f$ represents the effective superconducting gap for the f-electrons which is proximity induced not only by the surface $c$-electrons, but also by the bulk $c$-electrons. The inclusion of bulk bands in the Hamiltonian of Eq.(13) is computational too demanding to be implemented here, thus requiring the implementation of the effective gap $\Delta_f$ (similar to the approach employed in Ref. [6]).

To directly compare the theoretical DOS with the experimental $dI/dV$ shown in Fig. 3(b) of the main text on the pure Re and the La-Ce alloy surfaces, we note that the STS experiments are conducted in closed loop mode, implying that the distance between the tip and the surface is adjusted such that the total current between the tip and the surface for given set point voltage $V_s$ remains constant. As was shown in Ref. [7], this implies that in order to compare the theoretical DOS for the pure superconducting system with that in the presence of Kondo screening, the theoretical $N(\omega)$ needs to be renormalized according to

$$N_{ren}(\omega) = N(\omega) \left[ \int_0^{V_s} N(\omega) \right]^{-1} \tag{14}$$

The results for $N_{ren}(\omega)$ are shown in Fig. 3(f) of the main text, where we employed $V_s = 1\ meV$.

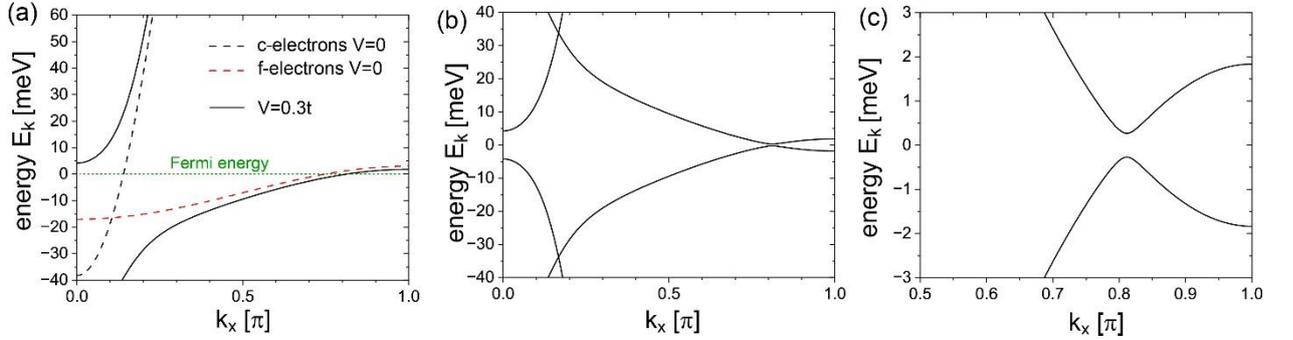

**Figure S1.**

(a) Dispersion in the normal state along the diagonal ($k_x = k_y$) in the Brillouin zone, for the unhybridized (V=0) and hybridized (V=0.3t) case. (b) Dispersion along the diagonal in the superconducting state with (c) being a zoom-in of (b).

## Section 3. Atomic-scale STM image of the ultrathin La-Ce alloy film in the vicinity of a Ce vacancy island with corresponding atomic structure model

To understand the atomic structure of the 2D La-Ce alloy, we have focused on vacancy defects in the La-Ce layer, which perturb the hexagonal lattice structure. We assume that these vacancy defects occur in the La-Ce layer due to an inhomogeneous concentration distribution of La and Ce atoms during the Ce deposition process. Additionally, the bonding strength between La and Ce could be relatively weak resulting in the local evaporation of Ce atoms into the vacuum during the annealing process. In Figure S2(a), a vacancy island is visible in the middle of the La-Ce layer. In the vicinity of the vacancy island, we observe an extended defect in the La-Ce layer with a slightly darker contrast. Figure S2(b) shows a zoomed-in STM image (a dotted square in Fig. S2(a)) focusing on the extended defect. A trimer-like structure is visible at the vacancy sites of the lattice formed by the bright protrusions. Based on the trimer-like structure as revealed by STM and the consideration of the deposited amount of La and Ce, we arrive at an atomic structure model as shown in Fig. 1(b), which basically corresponds to an ordered hexagonal array of $La_3Ce$ complexes. This model is consistent with the observed asymmetric double-peak resonance in the differential tunneling conductance spectra measured on the 2D La-Ce layer, reflecting that a large portion of electrons from the STM tip tunnel into the localized *f*-orbital localized at the Ce atoms being next to the tip apex.

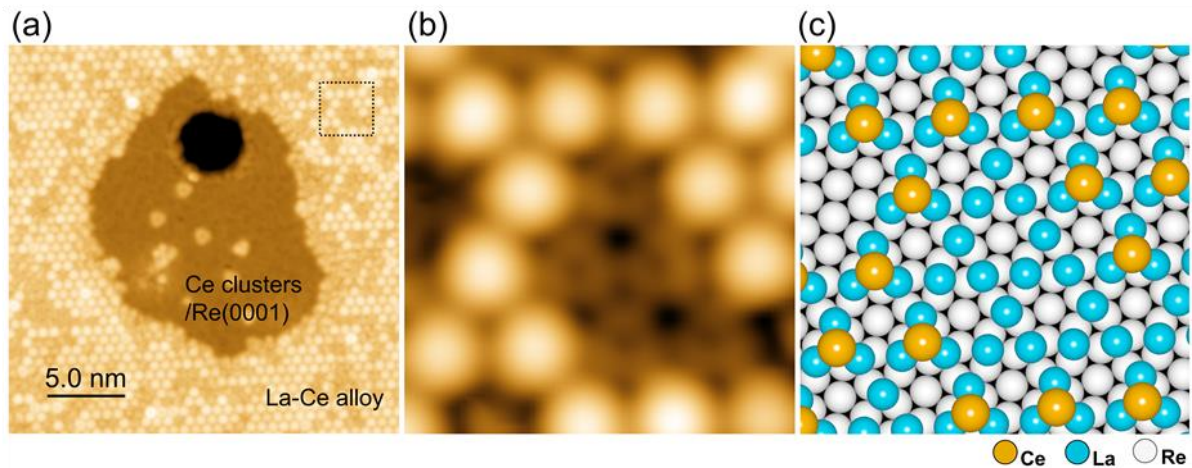

**Figure S2.**

(a) Constant-current STM topography image (25 × 25 nm$^2$) showing a vacancy island of the 2D La-Ce alloy at T = 1.7 K (1.0 nA/1.5 mV). The vacancy island shows the same characteristics as the terrace of the bare Re(0001) surface as shown in Fig. 1. (b) A zoomed-in STM image of the marked area (dotted square in (a)) revealing trimer-like structures at Ce-vacancy sites (25 × 25 nm$^2$). Holes shown as dark depressions are visible besides the trimer-like structures. (c) Atomic structure model based on the STM image in (b). Note that the trimer-like structures are formed by La atoms being in direct contact with the Re(0001) substrate. Each La trimer binds one Ce atom, thereby forming a La$_3$Ce complex. These La$_3$Ce complexes form an ordered hexagonal lattice on Re(0001).

*References*

[1]   S. Ouazi, T. Pohlmann, A. Kubetzka, K. von Bergmann, and R. Wiesendanger, Surf. Sci. **630**, 280 (2014).

[2]   D. K. Morr, Rep. Prog. Phys. **80**, 014502 (2017).

[3]   P. Coleman, Phys. Rev. B **29**, 3035 (1984).

[4]   D. M. Newns and N. Read, Adv. Phys. **36**, 799 (1987).

[5]   G. Kotliar and J. Liu, Phys. Rev. B **38**, 5142 (1988).

[6]   A. Palacio-Morales, E. Mascot, S. Cocklin, H. Kim, S. Rachel, D. K. Morr, and R. Wiesendanger, Sci. Adv. **5**, eaav6600 (2019).

[7]   J. Figgins, L. S. Mattos, W. Mar, Y.-T. Chen, H. C. Manoharan, and D. K. Morr, Nat. Commun. **10**, 5588 (2019).